\documentclass[conference]{IEEEtran}
\IEEEoverridecommandlockouts
\usepackage{graphicx}
\usepackage{graphicx}
\usepackage{multirow}
\usepackage{makecell}
\usepackage{amsmath}
\usepackage{amssymb} 
\usepackage{booktabs}
\usepackage{xurl}
\newcounter{algorithm}
\title{When to Test, Repair, or Regenerate: Decision-Guided Efficient RTL Generation}
\author{
\IEEEauthorblockN{
\textbf{Feiyang Pan}$^{1,\dagger}$,
\textbf{Shenghe Zheng}$^{2,\dagger}$,
\textbf{Pengfei Xiao}$^{3}$,
\textbf{Chunyan Yin}$^{1}$,
\textbf{Guangbin Dou}$^{1,*}$
}
\IEEEauthorblockA{
$^1$ Southeast University \\
$^2$ The Hong Kong University of Science and Technology \\
$^3$ University of Sheffield \\
\texttt{230238437@seu.edu.cn} \\
\textit{$^*$ Corresponding author: gdou@seu.edu.cn;
$^\dagger$ Equal contribution.}
}
}
\begin{document}
\maketitle
\raggedbottom
\begin{abstract}
Large language models (LLMs) can generate register-transfer-level
(RTL) designs from natural-language specifications.
To improve functional correctness, many RTL agents use simulation
feedback to guide refinement. However, without a reference
testbench, generating one for every initial implementation incurs
inference cost even when the RTL is correct.
We therefore introduce an adaptive harness that uses a specialized
decision model to orchestrate testing, refinement, and candidate
selection. It accepts high-scoring initial implementations and
generates testbenches for the rest. When further search is needed,
the initial score guides repair--regeneration allocation, while
defect rankings and simulation feedback guide repair.
For final selection, the harness prioritizes test-passing candidates
and ranks them using decision-model scores.
Across three benchmarks, the harness with DeepSeek improves mean
Pass@1 by 5.4--7.1 percentage points over initial implementations
and reduces aggregate estimated inference cost by 11.0--14.3\%
relative to a testbench-guided repair baseline with the same
candidate cap.

\end{abstract}


\noindent\textbf{Project Page:}
\url{https://github.com/2002-Pan/Decision-Guided-RTL}

\section{Introduction}
\label{sec:introduction}
Large language models (LLMs) have been widely explored for
register-transfer-level (RTL) generation and have shown promise in
translating natural-language specifications into hardware
designs~\cite{RavindranRTLReview,YangVerilogSurvey}. Functional
correctness, however, remains a challenge. An implementation may
compile successfully yet update a register one cycle too early or
handle reset differently from the specification. Detecting such
discrepancies requires evidence about the behavior of the generated
design. How an RTL generation agent obtains and uses that evidence
is therefore central to the reliability of its output.


RTL generation agents use compiler and simulation feedback to repair
faulty implementations~\cite{RTLFixer,VeriAssist}. This feedback can
also guide a search across candidates, as in MAGE, which combines
testbench generation, candidate sampling, and checkpoint-based
debugging~\cite{MAGE}. What an agent can infer about a candidate
depends on the evidence available during generation. With an
executable reference implementation, COEVO can compare candidate
outputs against expected values to guide repair and
selection~\cite{COEVO}. EstCoder uses a different source of evidence:
it estimates functional correctness from the candidate before
simulation and uses that estimate to guide acceptance, correction,
or regeneration~\cite{EstCoder}.

When only a natural-language specification is available, the harness
must decide whether a candidate warrants the cost of generating a
testbench before it knows how that candidate will perform. This
decision matters when the generator already solves most tasks on its
first attempt. In our study, for example, both generators pass at least 74\% of tasks on each of the three benchmarks before any testbench-guided refinement.
Generating a testbench for every task therefore incurs an additional
model call even for many implementations that already pass the
official tests. The challenge is to identify enough of these
implementations to save inference cost while keeping erroneous
acceptances rare. For candidates that undergo testing, a failure
raises a second allocation question: how should the harness divide
its remaining attempts between repairing the current design and
generating independent alternatives?

To address both decisions, we introduce an adaptive RTL generation
harness that uses a specialized decision model to score a candidate
against its specification. The controller accepts a high-scoring
initial RTL implementation as soon as it compiles. Otherwise, it
generates a testbench and returns the initial RTL if it passes.
When further search is needed, the initial score determines the mix
of repair and independent regeneration: intermediate scores favor
repair, while low scores favor regeneration. Repair uses available
testbench feedback and the model's two highest-ranked defect
categories; independent regeneration uses the specification alone.
For final selection, the controller scores each compilable candidate
independently. It returns the highest-scoring candidate that passes
the generated testbench, or the highest-scoring compilable candidate if none passes.

Our contributions are threefold.

\begin{enumerate}
\item We introduce an adaptive RTL generation harness that makes
testbench generation and subsequent search conditional on candidate
quality, allocating inference budget across verification, repair,
and independent regeneration.

\item We develop a structured assessment mechanism that combines
specialized decision model judgments with generated-testbench
feedback. It provides a common correctness measure across candidates
and ranked defect hypotheses for refinement, supporting control
decisions from the public specification and generated artifacts.

\item Across three benchmarks, the adaptive harness improves mean
Pass@1 by 5.4--7.1 percentage points over single-shot and
reduces aggregate estimated inference cost by 11.0--14.3\% relative
to a testbench-guided repair baseline. We also demonstrate transfer of the early-stop mechanism to an existing RTL agent framework.
\end{enumerate}

\section{Background and Motivation}

\label{sec:background}

\subsection{RTL Generation and Evaluation}

\textbf{Tool feedback and reference designs.}
Beyond improving RTL generators through domain-specific training and
data augmentation~\cite{RTLCoder,OriGen}, prior work uses design tools
to evaluate and revise candidates during inference. RTLFixer and MEIC~\cite{RTLFixer,MEIC}
use tool feedback to revise faulty RTL.
VerilogCoder~\cite{VerilogCoder} traces signals in waveforms from a supplied testbench
to diagnose functional failures. MAGE~\cite{MAGE} combines
testbench generation, candidate sampling, and checkpoint-based
debugging, incorporating a golden testbench when available.
Reference designs can also guide search. COEVO~\cite{COEVO} uses outputs from a
benchmark reference implementation to score candidates and guide
repair. In EvolVE's~\cite{EvolVE} structured testbench configuration,
an executable reference model supplies the expected outputs used to
evaluate search nodes.

\textbf{Evaluation from the specification alone.}
When only a specification is supplied, the workflow must obtain its
own evidence of correctness. ChipMATE~\cite{ChipMATE} generates RTL and a Python
reference model with separate agents and cross-checks their
behavior. NoTB~\cite{NoTB} generates candidates from multiple
model families and uses formal equivalence to identify implementations
that agree. EstCoder~\cite{EstCoder} uses a static functional estimate to guide
candidate acceptance, correction, or regeneration. Our focus is on
the cost of obtaining verification evidence: when to generate a
testbench and how to allocate the remaining generation budget if
that test exposes a failure.

\textbf{Specialized decision models.}
Recent specialized decision models, such as Jev~\cite{Jev} and the OpenAI Decisions API~\cite{OpenAIDecisions},
are designed to evaluate supplied inputs against user-defined criteria.
Given an input and an evaluation question, they return a score
indicating how strongly the input supports a particular judgment.
In an RTL harness, this capability raises the possibility of using
the specification and candidate RTL to judge whether further
verification is worth its cost.

\subsection{Motivation}

\begin{figure}[t]
  \centering
  \includegraphics[width=\columnwidth]{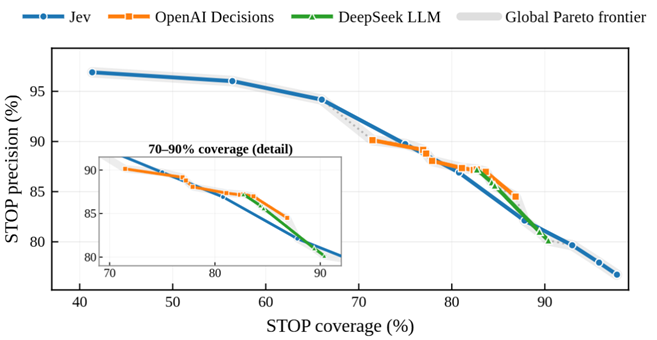}
  \par{\footnotesize (a) Early-stop precision versus coverage as the
acceptance threshold varies from 0.1 to 0.9.\par}

  \smallskip
  \includegraphics[width=\columnwidth]{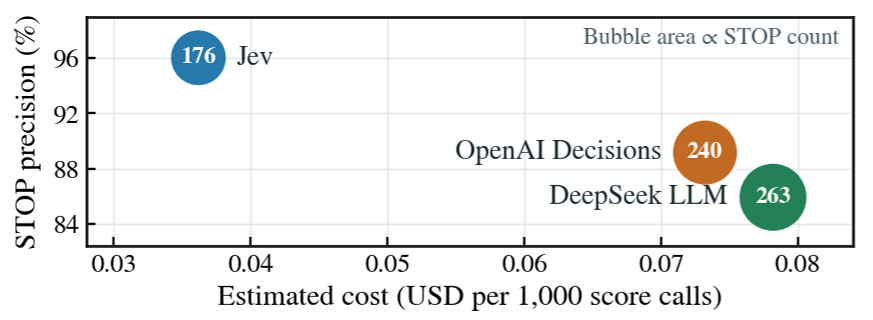}
  \par{\footnotesize (b) STOP precision and scoring cost at threshold 0.80.\par}
  \caption{Early-stop judgments on 312 evaluation instances, comprising
156 tasks from each of VerilogEval v1-Human and v2. Coverage includes
all instances, including initial compilation failures. In (b), bubble
labels give the number of accepted candidates.}
  \label{fig:early-stop}
\end{figure}

\textbf{Selective early stopping.}
We compare Jev, OpenAI Decisions, and a general-purpose DeepSeek
judge on the same initial RTL candidates. Each point in
Fig.~\ref{fig:early-stop}(a) corresponds to an acceptance threshold
between 0.1 and 0.9 applied to one judge's scores. Coverage is the
fraction of tasks accepted at that threshold, and precision is the
fraction of accepted implementations that pass the official tests.
The DeepSeek judge accepts a large fraction of candidates, and its
coverage changes little across several thresholds. The decision
models yield more distinct operating points: raising the threshold
progressively narrows the accepted set and increases its precision.
At the harness threshold of 0.80, both decision models achieve
higher precision at a lower estimated cost per scoring call than
the LLM judge (Fig.~\ref{fig:early-stop}(b)). These observations
motivate scoring the initial RTL before generating a testbench and
suggest a further use for the score: among candidates that subsequently
fail the generated testbench, different score ranges may warrant
different allocations of repair and regeneration attempts.

\textbf{Error categorization.}
In a preliminary study on MAEDA~\cite{MAEDA}, we evaluated whether
a decision model could identify specific types of errors in EDA
question answering. Our Jev-based evaluation pipeline achieved a
macro-F1 of 86.01\% across five error categories on 300 examples.
This result motivated us to explore a similar use of defect categories
in RTL repair, where each suggested category gives the repair model
a specific behavior to check against the specification.

\section{Method}
\label{sec:method}
\subsection{Harness Overview}

\begin{figure}[!htbp]
  \centering
  \includegraphics[width=\columnwidth]{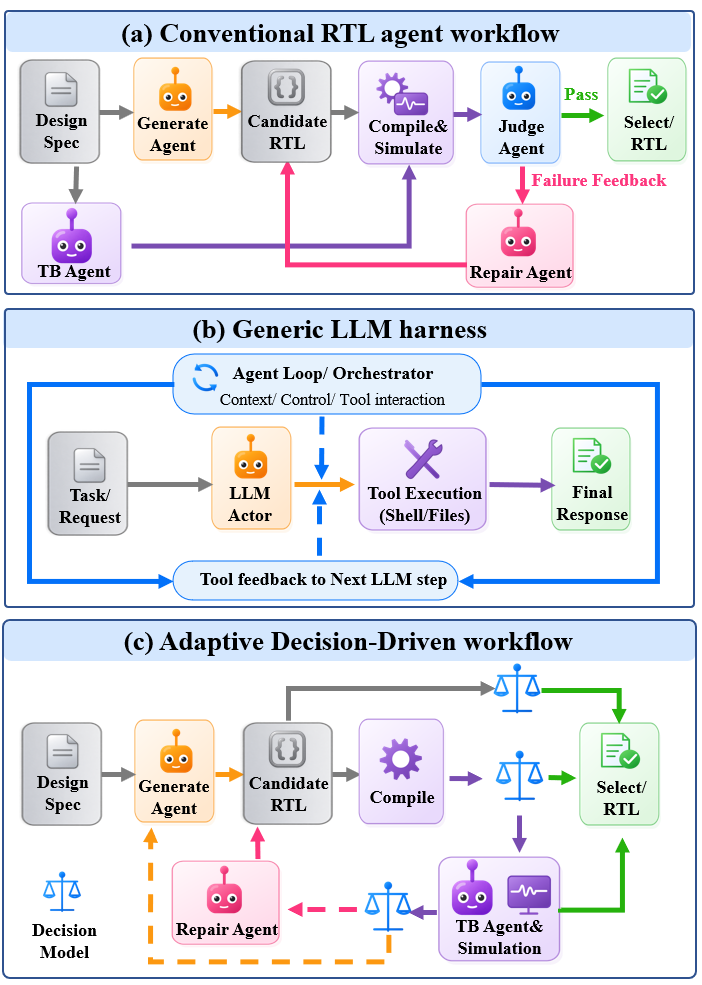}
 \caption{RTL generation and harness control.
  (a) Simulation-guided repair in a representative RTL agent.
  (b) Tool use and feedback in a general LLM harness.
  (c) Our adaptive harness uses decision-model scores for early
  stopping, budget allocation, and candidate selection.}
  \label{fig:workflow}
\end{figure}

In a representative RTL agent, simulation feedback guides acceptance
and repair (Fig.~\ref{fig:workflow}(a)). A general LLM harness also
controls when to obtain such feedback by deciding whether to invoke
a tool, request another model output, or return the current
result~\cite{MengAgentHarness,BarbasteHarnessEngineering}
(Fig.~\ref{fig:workflow}(b)). We bring this control into the RTL
workflow by pairing a specialized decision model with a deterministic
controller (Fig.~\ref{fig:workflow}(c)).

The harness first compiles the initial RTL, allowing up to two
compiler-guided repairs before declaring compilation failure.
Once the RTL compiles, the decision model scores it against the
public specification. The controller accepts a high-scoring
implementation immediately. Otherwise, an LLM generates a testbench
from the specification, and a passing initial implementation is
returned. When further search is needed, the controller reuses the
initial score to allocate attempts between repair and independent
regeneration. Both score ranges use both actions, with intermediate
scores favoring repair and low scores favoring regeneration.
When repair is requested, the decision model separately evaluates
predefined defect hypotheses. The controller aggregates their scores by category and passes the two highest-ranked categories to the repair model, which checks them against the specification and RTL alongside any available testbench feedback. Regeneration uses the specification
alone. For final selection, the controller independently scores each
compilable candidate and returns the highest-scoring one that passes
the generated testbench. If none passes, it returns the highest-scoring
candidate.

\subsection{Decision Model Interface}
The harness uses a specialized evaluator (Jev in our base
configuration) for two queries. Both take as input only the public
specification \(x\) and a compilable RTL candidate \(r\). The
correctness query scores the candidate as a whole; the diagnostic
query scores individual defect hypotheses when the controller
requests a repair.

\begin{figure}[!htbp]
  \centering
  \includegraphics[width=\columnwidth]{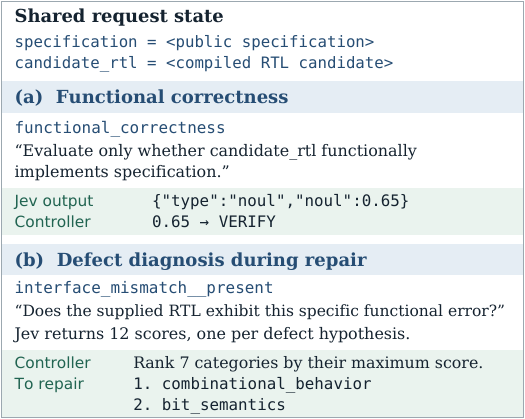}
  \caption{Decision model interface for RTL evaluation.
  (a) A correctness query returns a scalar score used by the
  controller for routing and candidate selection.
  (b) A diagnostic query scores defect hypotheses, which the
  controller aggregates into a ranked pair of category labels
  for repair. Example scores and rankings are illustrative.}
  \label{fig:jev-interface}
\end{figure}

\textbf{Correctness scoring.} Given a public specification \(x\) and a candidate RTL implementation \(r\) that compiles successfully, the harness queries decision model with a structured question asking whether \(r\) implements the behavior specified by \(x\) (Fig.~\ref{fig:jev-interface}(a)). The evaluator receives only \(x\) and \(r\). Its response is a scalar correctness score \(s(x,r)\in[0,1]\), where a higher value indicates stronger support for functional correctness. A deterministic controller uses the initial candidate's score to select a routing branch. When the harness generates additional candidates, it scores each compilable implementation independently using the same question and uses these scores in final candidate selection.

\textbf{Defect diagnosis.} When the harness requests a targeted repair, it submits the same specification and candidate to decision model with 12 questions covering seven defect categories (Fig.~\ref{fig:jev-interface}(b)). For example, three hypotheses check reset polarity, timing, and value. They can identify different discrepancies with the specification, yet all contribute to the same clock/reset category. A higher response score \(d_h\) indicates stronger evidence for the hypothesized defect. Let \(\mathcal{H}_c\) denote the hypotheses in category \(c\). The controller assigns the category a score
\begin{equation}
  D_c = \max_{h\in\mathcal{H}_c} d_h .
  \label{eq:defect-category-score}
\end{equation}
It ranks the categories by \(D_c\) and passes the two highest-ranked category names to the repair model in descending order. The repair model treats these labels as hypotheses to check against the specification and current RTL, using available feedback from the generated testbench to guide any correction.


\subsection{Adaptive Routing and Candidate Selection}
\label{sec:routing}

\textbf{Early acceptance.}
Let \(s_0=s(x,r_0)\) denote the correctness score of the initial
compilable RTL candidate \(r_0\). The controller returns \(r_0\)
immediately if \(s_0 \geq 0.80\) (Algorithm~\ref{alg:routing}).
For every lower score, the harness generates a testbench \(T\)
from the public specification \(x\) and tests \(r_0\). A pass
returns \(r_0\) regardless of its score. The distinction between
intermediate and low scores affects generation only after a valid
test failure.

\begin{figure}[t]
  \centering
  \includegraphics[width=\columnwidth]{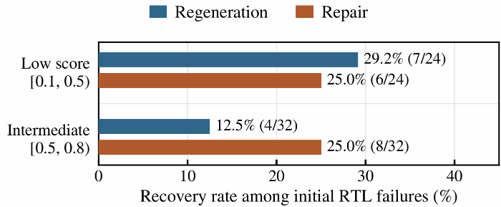}
  \caption{Recovery of initial RTL failures by repair and independent
  regeneration. Each label gives the number of recoveries over the
  number of observed failures in that score group.}
  \label{fig:routing-recovery}
\end{figure}

\textbf{Routing after a test failure.}
Figure~\ref{fig:routing-recovery} summarizes an exploratory analysis
of 56 initial RTL failures. At intermediate scores, repair recovered
8 of 32 failures and regeneration recovered 4. At low scores, the
corresponding counts were 6 and 7 of 24. We therefore retain both
actions in both branches and use the initial score to allocate the
third attempt. Section~IV evaluates the resulting policy end to end.

\noindent\begin{minipage}{\columnwidth}
\refstepcounter{algorithm}\label{alg:routing}
\smallskip\hrule\smallskip
{\small\textbf{Algorithm \thealgorithm}\quad
Adaptive routing and RTL selection\par}
\smallskip\hrule\smallskip
{\footnotesize
\textbf{Input}\quad Specification \(x\), compilable initial RTL \(r_0\),
score-based policy \(\pi_{\mathrm{inv}}\) for an invalid testbench\par
\smallskip
\begin{tabular}{@{}r@{\hspace{.55em}}p{\dimexpr\linewidth-2.2em\relax}@{}}
1  & \(s_0 \gets s(x,r_0)\), \(C \gets \{r_0\}\) \\
2  & \textbf{if} \(s_0 \geq 0.80\) \textbf{then return} \(r_0\) \\
3  & Generate testbench \(T\) from \(x\); let \(v \gets T(r_0)\) \\
4  & \textbf{if} \(v=\mathrm{PASS}\) \textbf{then return} \(r_0\) \\

5  & \textbf{if} \(v=\mathrm{FAIL}\) \textbf{then} \\
6  & \quad Try one repair and one independent regeneration \\
7  & \quad \textbf{if} \(s_0 < 0.50\) \textbf{then} \\
8  & \qquad Try a second independent regeneration \\
9  & \quad \textbf{else if} every candidate in \(C\) fails \(T\)
          and \(\max_{r\in C}s(x,r)<0.80\) \textbf{then} \\
10 & \qquad Try a second repair \\

11 & \textbf{else if} \(v=\mathrm{INVALID}\) \textbf{then} \\
12 & \quad \textbf{while} candidate budget remains \textbf{do} \\
13 & \qquad \(a \gets \pi_{\mathrm{inv}}(s_0,C)\) \\
14 & \qquad \textbf{if} \(a=\mathrm{STOP}\) \textbf{then break} \\
15 & \qquad Try the repair or regeneration specified by \(a\) \\

16 & \(P \gets \{r\in C : T \text{ yields a valid pass for } r\}\) \\
17 & \textbf{if} \(P\neq\varnothing\) \textbf{then return}
     \(\arg\max_{r\in P}s(x,r)\) \\
18 & \textbf{return} \(\arg\max_{r\in C}s(x,r)\) \\
\end{tabular}\par}
\smallskip\hrule\smallskip
{\footnotesize
Each attempt consumes one budget slot, up to three beyond \(r_0\).
Each compilable candidate is scored and added to \(C\).
Ties favor the earlier candidate.\par}

\end{minipage}
\par\addvspace{0.6\baselineskip}
A generated testbench yields \textsc{Pass} or \textsc{Fail} when
the simulation produces a verdict recognized by the harness.
We use \textsc{Invalid} when an attempted test yields no usable
verdict. Tasks accepted before testbench generation are recorded
separately. 
After a valid test failure, both branches retain \(r_0\) and allow
up to three additional candidates. Each branch first generates one
targeted repair using testbench feedback and the two highest-ranked
defect categories, and one independent implementation from \(x\)
alone. The recovery patterns in Fig.~\ref{fig:routing-recovery}
motivate how the controller allocates the third attempt.
For intermediate scores, where repair recovered more failures,
the controller reserves this attempt for a second repair. It uses
the attempt only if every compilable candidate fails \(T\) and the
highest correctness score among them remains below 0.80.
For low scores, where the two actions recovered similar numbers
of failures, the controller generates a second independent
implementation to explore an alternative to the initial design.

\textbf{Candidate selection.} The controller evaluates compilable candidates against the same generated testbench. Among candidates that pass, it returns the one with the highest score. If none passes or the testbench yields no valid verdict, the controller selects by score across all compilable candidates. Ties favor the earlier candidate. In the optional retry configuration, an invalid testbench triggers one replacement generated from the specification. If the replacement is valid, the controller reapplies the branch policy. A second invalid testbench leaves the base result in place. The official benchmark testbench is used only after the controller has selected its output.


\section{Experimental Evaluation}
\subsection{Experimental Setup}

We evaluate the harness on VerilogEval v1-Human~\cite{VerilogEval}, VerilogEval v2~\cite{VerilogEvalV2}
and RTLLM v2.0~\cite{RTLLM}. During generation and candidate
selection, the harness has access only to the public specification,
its generated artifacts, and feedback from local tools. The official
benchmark testbench is used only to evaluate the final selection.
We measure functional Pass@1 at the system level. For a benchmark
with \(N\) tasks, let \(y_{i,j}=1\) if the final RTL selected for
task \(i\) in run \(j\) passes the official testbench, and
\(y_{i,j}=0\) otherwise. Tasks that exhaust the compilation-repair
budget receive a zero. The pass rate for run \(j\) is
\begin{equation}
  \mathrm{Pass@1}_{j} = \frac{1}{N}\sum_{i=1}^{N} y_{i,j}.
  \label{eq:system-pass-at-one}
\end{equation}
For each benchmark and configuration evaluated in this work, we
report the mean Pass@1 over three independent runs.

\textbf{Model configurations and comparison protocol.}
In our primary configuration, DeepSeek V4.1 Flash generates both
RTL and testbenches. In the second configuration, GPT-5.6 Sol
performs both tasks. Both configurations use Jev-1.13.0 for
correctness scoring and defect diagnosis. To measure the gain from
the harness for each generator, we compare the initial RTL, including
any repairs made during the compilation step, with the final
RTL selected by the harness. We evaluate both on the same tasks
using the same official testbenches.

For comparisons with the testbench-guided DeepSeek baseline, both
systems allow up to three additional RTL candidates per task. We
report mean Pass@1 and estimated inference cost per task. The cost
estimate includes all model calls for RTL generation and repair,
testbench generation, correctness scoring, defect diagnosis, and
candidate selection. We apply the model-specific prices in effect
on October~31, 2026~\cite{DeepSeekPricing,OpenAIPricing,JevPricing}. For DeepSeek, we apply peak and off-peak rates
separately to the same recorded usage.
\subsection{Functional Correctness}

\textbf{Baselines and search budgets.}
We report prior systems under their published configurations.
MAGE ranks sampled RTL candidates for debugging and allows up to
five iterations for syntax repair~\cite{MAGE}. VerilogCoder allows
up to 100 consecutive agent replies~\cite{VerilogCoder}. For EvolVE,
we report the 15-node MCTS configuration without structured testbench
generation~\cite{EvolVE}. REvolution and COEVO each generate 10
offspring per generation for up to 20 and 10 generations,
respectively~\cite{REvolution,COEVO}. For a controlled comparison,
we evaluate our harness against a testbench-guided DeepSeek baseline
with the same cap of three additional RTL candidates per task.

\textbf{VerilogEval.}
Table~\ref{tab:verilogeval-comparison} places our results alongside
published RTL generation systems. On v2, the adaptive harness with
GPT-5.6 Sol reaches 93.1\% Pass@1. MAGE and COEVO report 95.7\% and
97.5\%, respectively, under different inference settings: MAGE can
incorporate a golden testbench when available, while COEVO uses a
benchmark reference implementation to obtain expected outputs during
search. Our harness generates its testbench from
the public specification. To measure the gain from the harness, we
compare against single-shot generation with the same backbone.
Mean Pass@1 improves by 5.4 and 7.1 percentage points on v1-Human
and v2 with DeepSeek V4.1 Flash.

\textbf{RTLLM.}
On the RTLLM v2.0 benchmark, our adaptive harness raises mean
Pass@1 from 74.0\% to 80.7\% with DeepSeek and from 81.3\%
to 86.7\% with GPT-5.6 Sol (Table~\ref{tab:rtllm-comparison}),
both generators again benefit.
The results for GPT-5.4-mini are taken from COEVO.
Notably, despite having no access to golden reference
implementations, our approach achieves competitive functional
correctness compared with methods that leverage such references.

\begin{table}[t]
  \centering
  \caption{Functional correctness on VerilogEval, measured by
           Pass@1 (\%). Results for prior systems are taken from
           their respective papers.}
  \label{tab:verilogeval-comparison}
  \begingroup
  \footnotesize
  \setlength{\tabcolsep}{3pt}
  \renewcommand{\arraystretch}{1.12}

  \resizebox{\columnwidth}{!}{%
    \begin{tabular}{c|c|c|c|c}
      \hline
      \multirow{2}{*}{\textbf{System}}
        & \multirow{2}{*}{\makecell{\textbf{Golden Ref.}\\\textbf{Access}}}
        & \multirow{2}{*}{\textbf{LLM Model}}
        & \multicolumn{2}{c}{\textbf{VerilogEval}} \\
      \cline{4-5}
        & & & \textbf{v1-Human} & \textbf{v2} \\
      \hline

      VeriAssist~\cite{VeriAssist}
        & Yes & GPT-4 & 50.5 & N/A \\
      \hline
      VerilogCoder~\cite{VerilogCoder}
        & Yes & GPT-4 Turbo & N/A & 94.2 \\
      \hline
      MAGE~\cite{MAGE}
        & Yes$^{\dagger}$
        & \makecell{Claude 3.5 Sonnet\\2024-10-22}
        & 94.8 & 95.7 \\
      \hline
      COEVO~\cite{COEVO}
        & Yes & GPT-5.4-mini & N/A & 97.5 \\
      \hline
      EvolVE~\cite{EvolVE}
        & NR & DeepSeek-R1 & N/A & 94.2 \\
      \hline
      REvolution~\cite{REvolution}
        & Yes & DeepSeek-V3 & N/A & 95.5 \\
      \hline\hline

      \multirow{6}{*}{Generic LLM}
        & No & GPT-4o & 51.3 & N/A \\
      \cline{2-5}
        & No & \makecell{Claude 3.5 Sonnet\\2024-06-20}
        & 60.3 & N/A \\
      \cline{2-5}
        & No & \makecell{Claude 3.5 Sonnet\\2024-10-22}
        & 75.0 & 72.4 \\
      \cline{2-5}
        & No & GPT-5.4-mini & N/A & 66.8 \\
      \cline{2-5}
        & No & \makecell{DeepSeek V4.1\\Flash$^{\ddagger}$}
        & 78.4 & 75.2 \\
      \cline{2-5}
        & No & GPT-5.6 Sol$^{\ddagger}$ & 90.8 & 90.3 \\
      \hline

      \multirow{2}{*}{\makecell{RTL-specific\\LLM}}
        & No & ITERTL~\cite{ITERTL} & 51.9 & N/A \\
      \cline{2-5}
        & No & CodeV~\cite{CodeV} & 53.2 & N/A \\
      \hline
      OriGen~\cite{OriGen}
        & No & \makecell{DeepSeek-Coder-7B\\+ LoRA}
        & 54.4 & N/A \\
      \hline
      AutoVCoder~\cite{AutoVCoder}
        & No & CodeQwen1.5-7B & 48.5 & N/A \\
      \hline
      AIvril~\cite{AIvril}
        & No & Claude 3.5 Sonnet & 64.7 & N/A \\
      \hline

      \multirow{2}{*}{\makecell{\textbf{Adaptive harness}\\\textbf{(ours)}}}
        & No & \makecell{DeepSeek V4.1\\Flash}
        & 83.8 & 82.3 \\
      \cline{2-5}
        & No & GPT-5.6 Sol & 93.4 & 93.1 \\
      \hline
    \end{tabular}%
  }

  \par\smallskip
  \begin{minipage}{\columnwidth}
    \footnotesize
    NR denotes unspecified reference access for the reported run.
    $^{\dagger}$MAGE uses the golden testbench when available.
    $^{\ddagger}$Single-shot results measured in this work.
  \end{minipage}
  \endgroup
\end{table}
\begin{table}[t]
  \centering
  \caption{Functional correctness on RTLLM. Results for prior
         systems are taken from their respective papers.}
  \label{tab:rtllm-comparison}
  \begingroup
  \footnotesize
  \setlength{\tabcolsep}{3pt}
  \renewcommand{\arraystretch}{1.12}

  \resizebox{\columnwidth}{!}{%
    \begin{tabular}{c|c|c|c}
      \hline
      \textbf{System}
        & \makecell{\textbf{Golden Ref.}\\\textbf{Access}}
        & \textbf{LLM Model}
        & \makecell{\textbf{RTLLM}\\\textbf{Pass@1 (\%)}} \\
      \hline

      COEVO\cite{COEVO}
        & Yes & GPT-5.4-mini & 94.5 \\
      \hline
      EvolVE\cite{EvolVE}
        & NR & DeepSeek-R1 & 82.0 \\
      \hline
      REvolution\cite{REvolution}
        & Yes & DeepSeek-V3 & 88.0 \\
      \hline\hline

      \multirow{3}{*}{Generic LLM}
        & No & GPT-5.4-mini & 64.4 \\
      \cline{2-4}
        & No & \makecell{DeepSeek V4.1\\Flash} & 74.0 \\
      \cline{2-4}
        & No & GPT-5.6 Sol & 81.3 \\
      \hline

      \multirow{2}{*}{\makecell{\textbf{Adaptive harness}\\\textbf{(ours)}}}
        & No & \makecell{DeepSeek V4.1\\Flash} & 80.7 \\
      \cline{2-4}
        & No & GPT-5.6 Sol & 86.7 \\
      \hline
    \end{tabular}%
  }

  \par\smallskip
  \begin{minipage}{\columnwidth}
    \footnotesize
    NR denotes unspecified reference access for the reported run.
  \end{minipage}
  \endgroup
\end{table}

\textbf{Comparison with testbench-guided repair.}
We compare the adaptive harness with a testbench-guided
baseline that uses DeepSeek V4.1 Flash for RTL generation,
testbench generation, repair, and judging. Both systems allow up
to three additional RTL candidates per task beyond the initial
implementation. The baseline generates a testbench from the public
specification and tests the initial RTL on every task. If the RTL
passes, the baseline returns it immediately. If a valid testbench
reports a failure, DeepSeek uses the simulation feedback to revise
the current RTL and tests the revision against the same testbench.
This loop stops at the first pass or after three repair attempts.
If the generated testbench yields no valid verdict, the baseline
generates three independent candidates and uses DeepSeek to select
one. Under the same candidate cap, our harness uses the decision
score to determine whether to spend the remaining budget on further
repair or independent regeneration.

\begin{figure}[t]
  \centering
  \includegraphics[width=\columnwidth]{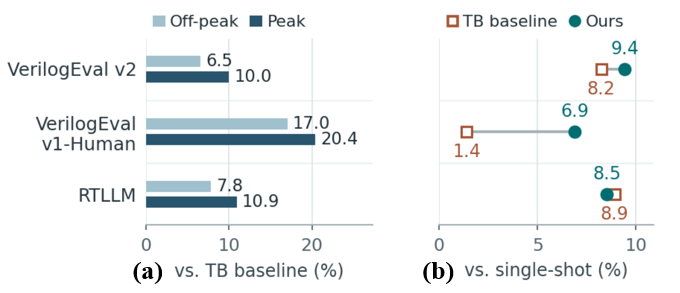}
  \caption{(a) Estimated inference cost
  savings of the adaptive harness relative to testbench-guided
  repair. (b) Relative Pass@1 gains of both methods over the
  initial DeepSeek implementations.}
  \label{fig:cost-pass-gain}
\end{figure}

Figure~\ref{fig:cost-pass-gain} shows how the two policies use this
budget in practice. The adaptive harness reduces estimated inference
cost on all three benchmarks. Relative to the testbench-guided
baseline, it saves 11.0\% in aggregate under off-peak pricing and
14.3\% under peak pricing. Its mean Pass@1 is 4.3 \%
higher on VerilogEval v1-Human and 0.9 points higher on VerilogEval
v2. On RTLLM, the harness achieves the cost reduction with a
0.3-point decrease in mean Pass@1.

\subsection{Decision Model and Routing Ablations}
On VerilogEval~v2, we vary the policy used after a test failure,
the diagnostic feedback supplied to the repair model, and the choice
of evaluator. Table~\ref{tab:decision-ablations} reports mean Pass@1,
early-stop rate, and estimated inference cost per task. 

\begin{table}[t]
  \centering
  \caption{Decision model and routing ablations on VerilogEval~v2
           with DeepSeek V4.1 Flash as the RTL generator.}
  \label{tab:decision-ablations}
  \begingroup
  \small
  \setlength{\tabcolsep}{2.5pt}
  \renewcommand{\arraystretch}{1.05}

  \begin{tabular*}{\columnwidth}{@{\extracolsep{\fill}}lrrr@{}}
    \toprule
    \textbf{Variant}
      & \makecell[r]{\textbf{Pass@1}\\(\%)}
      & \makecell[r]{\textbf{Early stop}\\(\%)}
      & \makecell[r]{\textbf{Cost}\\(\$/task)} \\
    \midrule
    \textbf{Full harness (Jev)}
      & \textbf{82.3} & 58.1 & 0.00123 \\
    \midrule
    \multicolumn{4}{@{}l}{Policy after test failure} \\
    \hspace{0.7em}Regeneration only
      & 79.2 & 58.1 & 0.00109 \\
    \hspace{0.7em}Repair only
      & 80.6 & 58.1 & 0.00134 \\
    \hspace{0.7em}2 repairs + 1 regeneration
      & 81.0 & 58.1 & 0.00126 \\
    \hspace{0.7em}1 repair + 2 regenerations
      & 80.3 & 58.1 & 0.00114 \\
    \midrule
    \addlinespace[3pt]
    \multicolumn{4}{@{}l}{Repair guidance} \\
    \hspace{0.7em}Without Top-2 diagnosis
      & 81.0 & 58.1 & 0.00119 \\
      \midrule
    \addlinespace[3pt]
    \multicolumn{4}{@{}l}{Evaluator replacement} \\
    \hspace{0.7em}OpenAI decision model
      & 80.8 & 72.4 & 0.00117 \\
    \hspace{0.7em}DeepSeek LLM judge
      & 77.6 & 78.8 & 0.00115 \\
    \bottomrule
  \end{tabular*}

  \par\smallskip
  \begin{minipage}{\columnwidth}
    \footnotesize
    
All rows report three-run means.

  \end{minipage}
  \endgroup
\end{table}

\textbf{Routing and repair guidance.}
We compare score-dependent routing with regeneration only, repair
only, and two fixed repair/regeneration mixtures under the same
early-stop rule and cap of three additional candidates. Adaptive
routing achieves 82.3\% mean Pass@1, outperforming regeneration only
by 3.1 percentage points and repair only by 1.7 points, with lower
cost than repair only. The fixed policies use two repairs and one regeneration, or vice versa. Adaptive routing improves Pass@1 over these mixtures by
1.3 and 2.0 percentage points, respectively, with a per-task cost
between those of the two fixed policies.

To assess the contribution of defect guidance, we keep adaptive
routing and testbench feedback but remove the two highest-ranked
defect categories from the repair prompt. Mean Pass@1 falls from
82.3\% to 81.0\%, while estimated cost decreases by \$0.00004 per
task.

\textbf{Evaluator choice.}
Replacing Jev with the OpenAI decision model or the DeepSeek LLM
judge increases early-stop coverage from 58.1\% to 72.4\% and
78.8\%, respectively. Both replacements reduce estimated cost per
task, while mean Pass@1 falls by 1.5 and 4.7 percentage points,
respectively. Under the routing policy evaluated here, Jev achieves
the highest end-to-end pass rate of the three evaluators.

\textbf{Budget scaling.}
We increase the cap \(N\) on additional RTL candidates from three
to six on VerilogEval~v2 while retaining the score-dependent routing
policy. After a valid test failure, the intermediate-score branch
allows one independent regeneration and up to \(N-1\) targeted
repairs. The low-score branch allows one targeted repair and up to
\(N-1\) independent regenerations. Thus, each increase in \(N\)
adds a repair opportunity for intermediate scores and a regeneration
opportunity for low scores.

Table~\ref{tab:budget-scaling} shows that increasing \(N\) from
three to six raises mean Pass@1 from 82.3\% to 83.3\%, at an
average of 0.65 additional actor calls per task relative to \(N=3\).
We use \(N=3\) as the default in the main experiments.


\begin{table}[t]
  \centering
  \caption{Budget scaling on VerilogEval~v2. Pass@1 is averaged
           over three runs. Extra actor calls are measured relative
           to the default budget of three additional RTL candidates.}
  \label{tab:budget-scaling}
  \begingroup
  \small
  \setlength{\tabcolsep}{3pt}
  \renewcommand{\arraystretch}{1.08}

  \begin{tabular*}{\columnwidth}{@{\extracolsep{\fill}}lrrrr@{}}
    \toprule
    \textbf{Candidate budget}
      & \textbf{3} & \textbf{4} & \textbf{5} & \textbf{6} \\
    \midrule
    Mean Pass@1 (\%)      & 82.3  & 82.7  & 82.9  & 83.3  \\
    Extra actor calls/task & 0.000 & 0.218 & 0.436 & 0.654 \\
    \bottomrule
  \end{tabular*}
  \endgroup
\end{table}

\textbf{Transfer to MAGE.}
We evaluate the early-stop gate in Stable-Lab's open-source MAGE
implementation~\cite{MAGE}. Both configurations use DeepSeek and
retain MAGE's original agents and prompts. The reference RTL is
withheld, and the official testbench is used only for final
evaluation. We insert the gate after MAGE has generated its
testbench and the initial RTL has passed syntax checks, but before
the first simulation. Jev scores the initial RTL against the public
specification. If the score is at least 0.80, MAGE returns the
initial RTL, skipping simulation and any subsequent calls for
diagnosis or refinement. Otherwise, MAGE continues its standard
workflow. On VerilogEval~v2, adding the gate reduces mean DeepSeek
calls per task from 3.327 to 2.833 (14.8\%), with a
1.1-percentage-point decrease in Pass@1.

\textbf{Relation to static functional estimation.}
EstCoder uses static functional estimates to guide acceptance,
correction, or regeneration~\cite{EstCoder}. Our harness extends
this decision process to selective acquisition of simulation
evidence: it determines when to generate a testbench, combines
test outcomes with static assessments to allocate further
attempts, and ranks candidates for final selection.
\section{Conclusion}
In this paper, we introduced an adaptive harness for generating RTL
from a natural-language specification. The harness uses a specialized
decision model to score each initial RTL implementation that compiles.
The score guides two decisions at different stages: whether to
accept the implementation before generating a testbench and, if the
generated testbench exposes a failure, how to divide the remaining
candidate budget between targeted repair and independent regeneration.

Across three benchmarks, the adaptive harness improved mean
Pass@1 by 5.4--7.1 percentage points over the initial DeepSeek
implementations. On 312 VerilogEval tasks, its early-stop criterion
selected 56.4\% of initial implementations for acceptance without
testbench generation, and 96.0\% of those accepted passed the official
tests. Compared with a testbench-guided repair baseline under the same
candidate cap, the harness reduced aggregate estimated inference cost
by 11.0--14.3\%. These results show how an RTL agent can reserve
verification and refinement effort for selected candidates while
improving the functional correctness of its final outputs.

\bibliographystyle{IEEEtran}
\bibliography{references}
\end{document}